\documentclass[twocolumn,pra,aps]{revtex4-1}
\usepackage{amsmath, amssymb, graphicx}
\usepackage{hyperref}
\usepackage{braket}
\usepackage{indentfirst}
\usepackage{subcaption}
\usepackage{CJKutf8}

\def\be{\begin{equation}}
\def\ee{\end{equation}}
\def\ba{\begin{align}}
\def\ea{\end{align}}
\def\sdg{Schr\"odinger~}

\begin{document}

\title{How Are Quantum Eigenfunctions of Hydrogen Atom Related To Its Classical Elliptic Orbits?}

\begin{CJK*}{UTF8}{gbsn}
\author{Yixuan Yin}
\affiliation{International Center for Quantum Materials, School of Physics, Peking University,  Beijing 100871, China}
\author{Tiantian Wang}
\affiliation{International Center for Quantum Materials, School of Physics, Peking University,  Beijing 100871, China}
\author{Biao Wu}
\email{wubiao@pku.edu.cn}
\affiliation{International Center for Quantum Materials, School of Physics, Peking University,  Beijing 100871, China}
\affiliation{Wilczek Quantum Center, Shanghai Institute for Advanced Studies, 
Shanghai 201315, China}
\affiliation{Hefei National Laboratory, Hefei 230088, China}
\date{\today}

\begin{abstract}
We show that a highly-excited energy eigenfunction $\psi_{nlm}(\vec{r})$ of hydrogen atom can 
be approximated as  an equal-weight superposition of classical elliptic orbits with energy $E_n$ and
angular momentum $L=\sqrt{l(l+1)}\hbar$, and  $z$ component of angular momentum $L_z=m\hbar$. 
This correspondence is established by comparing the quantum probability
distribution $|\psi_{nlm}(\vec{r})|^2$ and the classical probability
distribution $p_c(\vec{r})$ of an ensemble of such orbits.  This finding illustrates a general 
principle: in the semi-classical limit,  an energy eigenstate of a quantum system is 
in general reduced to a collection of classical orbits, rather than a single classical orbit. 
 
\end{abstract}	
\maketitle

\end{CJK*}

\section{Introduction}
\begin{figure}[ht]
	\centering
	\includegraphics[width=0.4\textwidth]{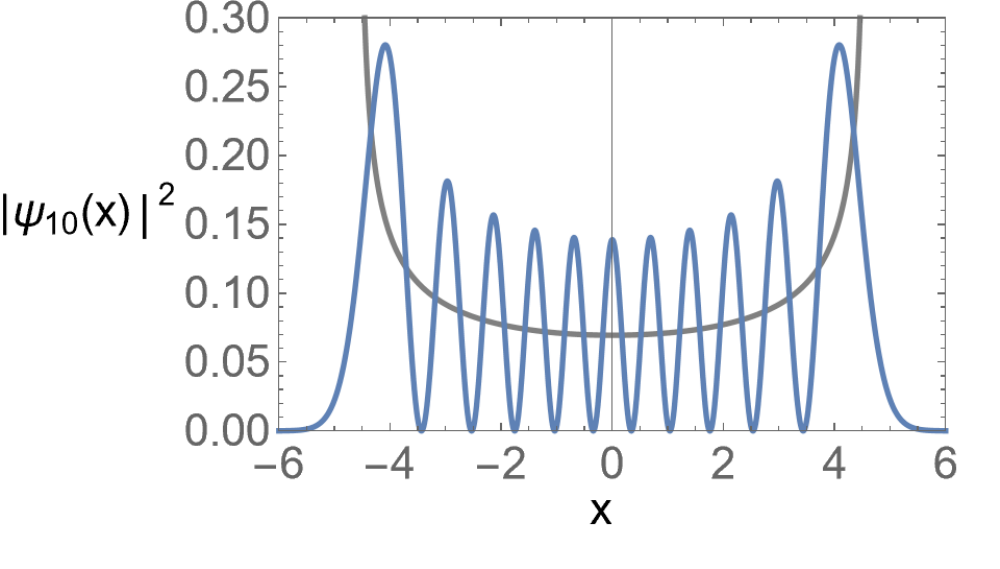}
	\caption{Probability density of a harmonic oscillator.  Blue line is the quantum probability density for
		the 10th energy eigenstate; the gray line is the corresponding classical probability density. 
		The unit of  $x$ is $\sqrt{\frac{m\omega}{\hbar}}$, where $m$ and $\omega$ are the mass and frequency of
		the oscillator, respectively.}
	\label{ho}
\end{figure}

Classical mechanics is an approximation of quantum mechanics in the limit of 
large quantum numbers or,  equivalently,  as the Planck constant $\hbar$ approaches zero. 
This quantum-classical correspondence
is often illustrated in textbooks with one dimensional harmonic oscillator\cite{Alber,Pauling}
by comparing the probability density $|\psi_n(x)|^2$ of an energy eigenfunction to
the  probability density of the corresponding classical orbit, as shown 
in Figure \ref{ho}. However, this example may be misleading, as it suggests a 
one-to-one  correspondence between a quantum energy eigenfunction and a single classical orbit. 
In reality, for systems with multiple degrees of freedom, a highly excited energy eigenfunction corresponds 
to an ensemble of classical orbits. More precisely, in the semiclassical limit $\hbar\rightarrow 0$, 
an energy eigenfunction is reduced to a superposition of many different classical orbits, which share 
the same energy but may differ in other dynamical variables. Such a superposition  
forms a distribution in classical phase space, which is invariant under classical evolution\cite{berry1,berry2,Zhang}. 

\begin{figure}[ht]
	\centering
	\begin{subfigure}[t]{0.3\textwidth}
		\centering
		\includegraphics[width=1\textwidth]{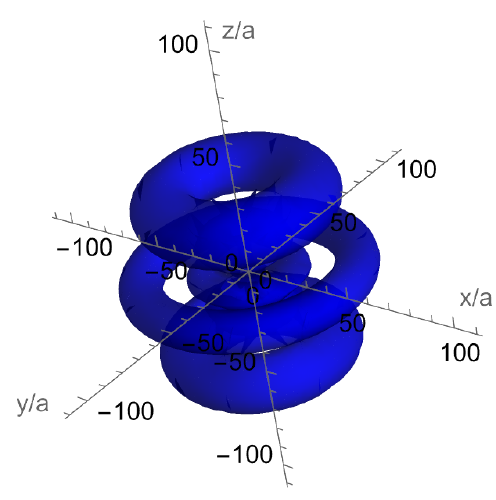}
	\end{subfigure}
	\quad
	\begin{subfigure}[t]{0.3\textwidth}
		\centering
		\includegraphics[width=1\textwidth]{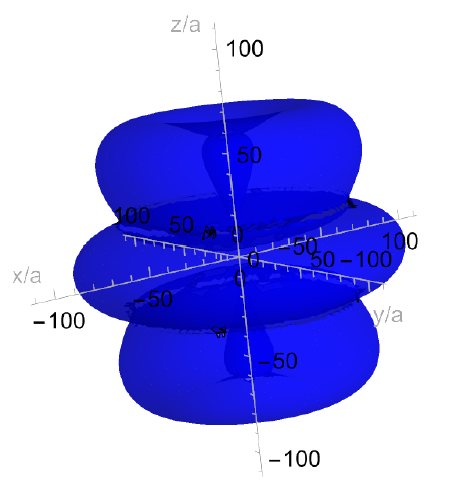}
	\end{subfigure}
	\quad
	\caption{Comparison of the quantum and classical probability densities: 
		(a) the surface where the quantum probability  is $|\psi_{nlm}|^2=10^{-5} \text{m}^3$ 
		with $n=6$, $l=4$, $m=2$; (b) 
		the surface where the classical probability divided by $\sin\theta$ is $10^{-5} \text{m}^3$. In the figure, $a$ is the Bohr radius.}
	\label{quan}
\end{figure}

In this work, we employ the hydrogen atom as a paradigmatic example to elucidate the general quantum-classical correspondence. Specifically, we demonstrate  that a highly excited eigenstate  $\psi_{n,l,m}(\vec{r})$ can be interpreted 
as an  equal-weight superposition of all possible classical elliptic orbits 
with energy $E=E_n$, angular momentum $L=\sqrt{l(l+1)}\hbar$, and $z$-component of angular momentum $L_z=m\hbar$.  
We are able to compute the probability density for the ensemble of  superposed elliptic orbits and compare it the quantum probability density $|\psi_{nlm}(\vec{r})|^2$. There is remarkable agreement as seen in Figure \ref{quan}. 
For hydrogen atom,  the necessity of this ensemble representation for 
an eigenfunction $\psi_{nlm}(\vec{r})$– as opposed to a single-orbit correspondence – 
reflects that the three good quantum number $n,l,m$ can not completely specify a single classical orbit.

In addition, the quantum-classical correspondence established here justifies the approximation of a quantum state of electron 
with classical orbits or classical ensemble distribution, which is often used in the study 
of ionization of atoms in strong fields\cite{Abrines,unclose,Cui,Liu}.

In this article, Section 2 introduces the quantum probability density as described in standard quantum physics textbooks. 
Section 3 derives the classical probability density, followed by a comparison of the quantum and classical results in Section 4. 
The overall findings are discussed in Section 5, and the Appendix addresses singularities 
with the quantum-classical correspondence established in this work.

\section{Quantum probability density}
The quantum solutions for the hydrogen atom are well-established. 
For completeness and to introduce notations, we briefly review them here. 
The  energy eigenfunction $\psi$ of the hydrogen atom satisfies  
the following time-independent \sdg equation:
\begin{equation}
	-\frac{\hbar^2}{2m_e}\nabla^2\psi(\vec{r})+V(r) \psi(\vec{r}) =E \psi(\vec{r})\,,
\end{equation}
where $m_e$ is the mass of the electron (the mass of proton is approximated as infinite) 
and $V(r)$ is the Coulomb potential  
\begin{equation}
	V(r)=-\frac{q^2}{4 \pi \epsilon_0 r}\,.
\end{equation}	
Here $q$ is the charge of the electron. The solutions of this equation are well-known and can be found in standard textbooks~\cite{Griffiths}.  
In the spherical coordinates, they can be expressed as:
\begin{equation}
	\psi_{nlm}(r,\theta,\phi)=R_{nl}(r) Y_l^m(\theta,\phi)\,,
\end{equation}
where
\begin{equation}
	\label{eq:R}
	R_{nl}(r)=\sqrt{\Big(\frac{2}{na}\Big)^3 \frac{(n-l-1)!}{2n(n+l)!}} 
	\Big(\frac{2r}{na}\Big)^l e^{-\frac{r}{na}} L_{n-l-1}^{2l+1}(\frac{2r}{na})\,,
\end{equation}
and 
\begin{equation}
	Y_l^m(\theta,\phi)=\sqrt{\frac{2l+1}{4\pi} \frac{(l-m)!}{(l+m)!}}e^{im\phi}P_l^m(\cos\theta)\,.
\end{equation}
Here $n$ is an integer, $l$ is an integer satisfying $0\leq l \leq n-1$, $m$ is an integer satisfying $-l \leq m\leq l$ and $a$ is the Bohr radius and is given by
\begin{equation}
	a=\frac{4 \pi \epsilon_0 \hbar^2}{m_e q^2}\,.
\end{equation}
For the eigenfunction $\psi_{nlm}$, the corresponding eigenenergy is
\begin{equation}
	E_{n}=-\frac{m_eq^4}{32\pi^2\epsilon_0^2\hbar^2n^2}=-\frac{\hbar^2}{2m_e a^2n^2}\,,
\end{equation}
the angular momentum is $L=\sqrt{l(l+1)} \hbar$, and its $z$ component  is $L_{z}=m \hbar$. 

In the WKB approximation, the angular momentum is chosen as $L=(l+1/2)\hbar$~\cite{Robnik}. 
Here we mostly consider the situation with large $l$, which means the difference 
between $\sqrt{l(l+1)}$ and $l+1/2$ could be neglected.

We are interested in the classical counterpart of 
the quantum probability density $|\psi_{nlm}(r,\theta,\phi)|^2$ 
in the semiclassical limit $\hbar\rightarrow 0$ or, equivalently, the limit of $n\rightarrow \infty$. 
For the one-dimensional  harmonic oscillator, in such a limit, 
the quantum probability distribution for an energy eigenfunction
corresponds to the  probability density of a single classical periodic orbit as shown in Fig. \ref{ho}. 
This  is actually true for any time-independent one-dimensional system: one eigenfunction corresponds to 
a single classical orbit.  

However, for a general system with two or three dimensions,  this assertion is not accurate. 
Consider the case of zero angular momentum, i.e., $l=0$. In this case,  
the eigenfunction $\psi_{n00}(r)$ of hydrogen atom is spherically symmetric. In contrast, the classical orbit
in this case is a straight line, bearing no resemblance to the quantum state $\psi_{n00}(r)$ at all. 

\begin{figure*}[htbp]
	\centering
	\includegraphics[width=0.9\linewidth]{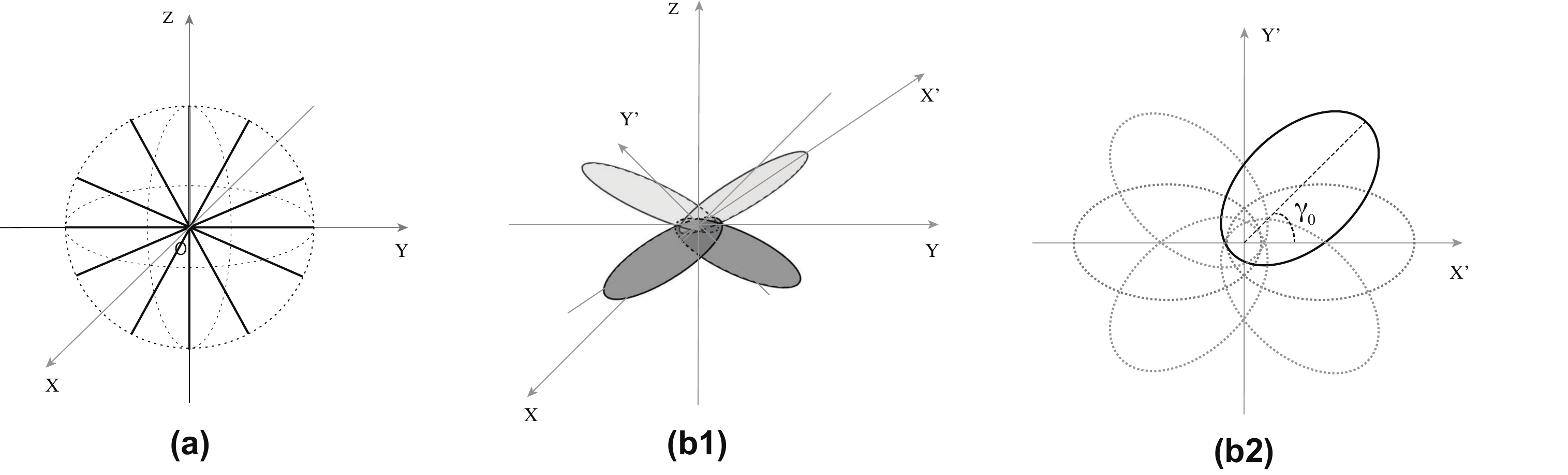}
	\caption{Classical orbits of hydrogen atom. The origin of the coordinates is the position of proton. 
		(a) In the special case of zero angular momentum $l=0$, the classical orbits are straight lines orientated 
		in all possible directions.  These orbits collectively fill a spherical region , 
		reflecting that the corresponding eigenstates 
		are $s$-waves. The  radius of this sphere is determined by the eigen-energy $E_n$.  
		(b) For non-zero angular momentum $l\neq 0$, 
		the classical orbits for a quantum eigenstate $\psi_{nlm}(\vec{r})$ are ellipses, which 
		can be divided into two groups: (1)
		the ellipses that can be transformed into each other rotating around the $z$ axis and reflecting off the $XY$ plane; 
		(2) all the ellipses that rotate  around the focus of proton in the orbital plane, 
		which is perpendicular to the angular momentum $\vec{L}$. For convenience,  
		the $X'Y'$ coordinate system is set up in the orbital  plane.  
		The shape and size of the ellipses are determined by quantum number $n$ and $l$, 
		the orientation of the ellipses relative to the $Z$ axis is determined by $l$ and $m$. }
	\label{fig:orbit}
\end{figure*}

To reconcile this discrepancy, we note that  the three good quantum numbers $n, l=m=0$ in the 
eigenfunction $\psi_{n00}(r)$ do not completely determine the classical orbit as 
the  direction of the classical orbit remains unspecified. 
As shown in Figure \ref{fig:orbit}(a),  all the straight lines inside a sphere, whose radius is determined by $E_n$, 
are possible classical orbits corresponding to the eigenfunction $\psi_{n00}(r)$. 

When the angular momentum is non-zero, $l\neq 0$, the classical orbit is elliptical. Similarly, these ellipses are not
completely specified by the three good quantum numbers $n,l,m$.  The size and shape of the ellipses
(the major and minor axes) are determined by $n,l$, while the angle between the ellipse plane and the $Z$ axis 
is determined by $l,m$.   As shown in Fig.\ref{fig:orbit}(b1), when an ellipse rotates around the $Z$ axis or gets reflected 
with respect to the $XY$ plane, its corresponding quantum numbers $n,l,m$ remain unchanged. 
Moreover, as shown in Fig.\ref{fig:orbit}(b2), if an ellipse rotates in its plane around the focus (where the proton is),  
its corresponding quantum numbers $n,l,m$ do not change, either. 
All these ellipses in Fig.\ref{fig:orbit}(b1, b2) correspond to the same $n,l,m$ and form an ensemble.

As we will see, the quantum eigenfunction $\psi_{nlm}(\vec{r})$ corresponds to such an ensemble classical orbits. 
We will compute 
the classical probability density $p_c(\vec{r})$ for such an ensemble, comparing it to the quantum 
quantum probability density $|\psi_{nlm}(\vec{r})|^2$. We find that in the semiclassical limit, 
these two densities agree with each other very well.  Specifically, we will compare 
the quantum radial probability density $p_q(r)=|R_{nl}(r)|^2$  and 
its angular probability density $p_q(\theta)=|Y_l^m(\theta,\phi)|^2$ to their classical 
counterparts,  respectively.

\section{Classical probability density}
\begin{figure}[ht]
	\centering
	\includegraphics[width=0.4\textwidth]{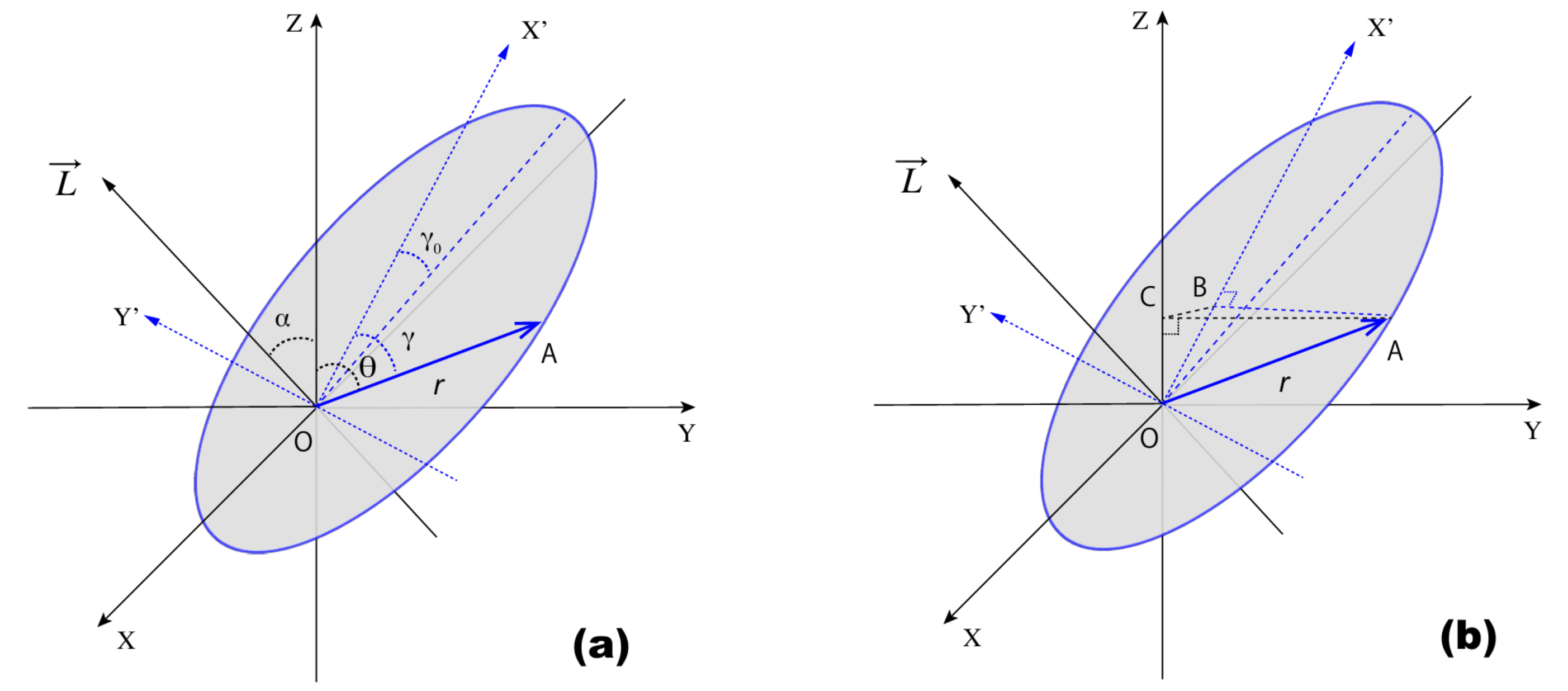}
	\caption{One elliptic orbit in a coordinate system where the origin is the position of proton. 
		$\vec{L}$ is the angular momentum vector.  A is a point on the elliptic orbit. 
		All the blue lines lie in the plane of the elliptic orbit. (a) $\alpha$ is
		the angle between $\vec{L}$ and the $Z$ axis, and  
		$\theta$ is the angle between OA and the $Z$ axis. 
		$X'Y'$ is a two-dimensional coordinate system in the  plane of elliptic orbit. 
		The $X'$ axis, $Z$ axis, and $\vec{L}$ lie in the same plane. $\gamma_0$ is the angle between 
		the major axis of the ellipse and the $X'$-axis, and $\gamma$ is the  angle between OA 
		and the $X'$ axis. (b) B is a point on the $X'$ axis, and C is a point on the $Z$ axis.  
		Their positions are chosen so that triangle ABC is 
		perpendicular to the $Z$ axis. }
	\label{fig:ellipse}
\end{figure}

For a typical elliptical orbit,  its standard form  is given by\cite{Goldstein}
\be
\label{eq:rp}
r(t)=\frac{p}{1-\epsilon\cos\gamma'(t)}\,,
\ee
where the parameters $p$ and $\epsilon$ are determined by the energy $E$ and 
angular momentum $L$ of the orbit. The angle $\gamma'=\gamma-\gamma_0$ is 
illustrated  in Figure \ref{fig:ellipse},  where we have set up a coordinate system for an elliptical orbit. 
The position of proton is the origin O of the axes and thus is a focus of the ellipse,  
A is the position of electron at time $t$,   $\vec{L}$ is the angular momentum vector 
that is perpendicular to the the orbital plane, and $\theta$  is angle between OA and the $Z$ axis. 
For convenience of discussion, we set up 
a two-dimensional coordinate system $X'Y'$ in the orbital plane, whose origin is also O. 
The $X'$ axis is set up such that it lies in the plane defined by $\vec{L}$ and the $z$ axis. 
$\gamma$ and $\gamma_0$  are the angles of OA and the major axis with respect to the $X'$ axis. 
We pick a point B on the $X'$ axis such that AB is perpendicular to the $X'$ axis; similarly, C is a point 
on the $Z$ axis such that AC is perpendicular to the $z$ axis. 

For a hydrogen atom in the eigenfunction $\psi_{nlm}(\vec{r})$,  
both its angular momentum and energy have definite values: 
$L=\sqrt{l(l+1)}\hbar$  and 
\begin{equation}
	\label{eq:en}
	E_{n}=-\frac{\hbar^2}{2m_e a^2n^2}=\frac{1}{2}m_ev^2-\frac{q^2}{4\pi\epsilon_0 r}\,.
\end{equation}
where $v$ is the electron's speed.  A classical orbit corresponding to this eigenfunction 
shares  the same energy $E_n$ and angular momentum $L$. Using the standard definition~\cite{Goldstein}, 
we can obtain the semilatus rectum $p=a\ell^2$ and the eccentricity
\be
\label{eccen}
\epsilon=\sqrt{1-\frac{\ell^2}{n^2}}\,,
\ee
where $\ell^2=l(l+1)$. For the eigenfunction  $\psi_{nlm}(\vec{r})$, the  $Z$ component  
of angular momentum is also definite with $L_z=m\hbar$.  
This value of $L_z$ fixes the angle $\alpha$ between the $Z$ axis and 
the angular momentum vector $\vec{L}$, as shown in Figure \ref{fig:ellipse}. 

As already mentioned above, these three quantities--$E_n$, $L$, and $L_z$, specified 
by the quantum numbers $n,l,m$, respectively--along with the position of the proton 
do not uniquely determine the elliptic orbits. Two additional parameters,  $L_x$ (or $L_y$) 
and $\gamma_0$, are needed. Shown in Figure \ref{fig:orbit}(b1) 
are four such orbits sharing identical $E_n$, $L$, and $L_z$ values with $\gamma_0=0$ but different $L_x$; 
the ellipses in Figure \ref{fig:orbit}(b2) have the same values of $E_n$, $L$, $L_z$, and $L_x$ with 
different $\gamma_0$. Due to the spherical symmetry of the hydrogen atom, it is natural to interpret 
the eigenfunction $\psi_{nlm}(\vec{r})$  as an equal-weight superposition 
of  all these classical orbits with the same $E_n$, $L$, and $L_z$ but different $L_x$ and $\gamma_0$. 
We will calculate the probability densities for this ensemble, 
and compare them to the  quantum counterparts, finding excellent agreement.

The elliptic orbit is  closed and periodic. For  a closed classical orbit $s(t)$ 
with period $T$ (i.e., $s(0)=s(T)$), the probability $p_c(s)$ of finding the classical particle
between $s(t)$ and $s(t+\delta t)$ is defined as 
\be 
p_c(s)\delta s\propto \frac{\delta t}{T}=\frac{\delta s}{vT}\,,
\ee 
where $\delta s=s(t+\delta t)-s(t)$ and $v$ is the speed of the particle at $s(t)$. 
It is evident that $p_c$ diverges at $s(t)$, where $v$ is zero. This approach has been employed to 
calculate  the classical probability density of a harmonic oscillator, 
as depicted by the gray line in Figure \ref{ho}, which diverges at the two end points
of the oscillations.  We will adopt this method to compute the classical radial 
probability density  $p_c(r)$ and angular probability density  $p_c(\theta)$, and compare
them to their quantum counterparts. 
All the classical orbits in the classical ensemble corresponding 
to the eigenfunction $\psi_{nlm}(\vec{r})$  have the same period, 
\be
T=2\pi \frac{m_e a^2n^3}{\hbar}\,,
\ee
which will be used in our calculations. 

\subsection{Radial probability density}
We first consider the special case of zero angular momentum, $l=0$, for two reasons: 
(1) to use this simplest case to illustrate how to compute the classical probability density; (2) this
special case cannot be obtained by taking the limit $l\rightarrow 0$ from the result of $l\neq 0$.  
In this case, as shown in Figure \ref{fig:orbit}(a), the orbits consist of all the straight lines through
the center inside a sphere, whose radius is determined by the energy $E_n$. 
We focus on one of the straight lines and use $r$, the distance to the center, 
as the coordinate for this one dimensional motion. So, the velocity is given by $v=dr/dt$. 
Using Eq.(\ref{eq:en}), we find that the time 
the electron spends between radial distance between $r$ and $r+dr$ is 
\begin{equation}
	dt=2\frac{dr}{v}=2\frac{m_edr}{\hbar \sqrt{\frac{2}{ar}-\frac{1}{n^2a^2}}}\,.
\end{equation}
Here the factor of  2 accounts for  the electron  passing through a specific 
radius $r$ twice per  period. Since the probability between $r$ and $r+dr$ is defined as $p_c(r)dr=dt/T$, 
we have the probability density $p_c(r)$ as 
\begin{equation}
	\label{l20}
	p_{c}(r)=\frac{1}{\pi  n^3 a\sqrt{\frac{2}{\tilde r}-\frac{1}{n^2}}}\,,
\end{equation}
where $\tilde{r}=r/a$. This classical density diverges at $\tilde{r}=2n^2$, 
the farthest point away from the center where the electron has zero velocity. 

Note that,  to be precise, for a single orbit, the above density 
should be written as $p_{c}(r)\delta(\theta-\theta_0)\delta(\phi-\phi_0)$, where 
$\theta_0$ and $\phi_0$  specify the direction of the motion. Since the density $p_{c}(r)$ is independent 
of the direction, the radial probability density for the ensemble of all the possible orbits with 
equal weight, i.e., the sphere in Fig. \ref{fig:orbit}(a), should be just $p_{c}(r)$. 
This shows that we can obtain the classical probability density by focusing just on a single orbit. 
Other cases are similar. As a result,  for brevity,  in the following computation, 
we will not distinguish the difference between the probability density for a single orbit and the density 
for the ensemble.  

We now consider the case where the  angular momentum is not zero ($l\neq 0$). 
In this case, the distance that the electron moves from $r$ to $r+dr$ is
\begin{equation}
	\label{eq:ds}
	ds=\sqrt{(dr)^2+(rd\gamma)^2}\,.
\end{equation}
Using Eq.(\ref{eq:rp}), we find that 
\begin{equation}
	ds=dr\sqrt{1+\frac{a^2 \ell^4}{r^2 \big[1-\frac{\ell^2}{n^2}-(1-\frac{a\ell^2}{r})^2\big]}}
\end{equation}
The time that the electron spends in the range from $r$ to $r+dr$ is $dt=\frac{ds}{v}$. 
Therefore, the probability density at $r$ is: 
\begin{equation}
	p_{c}(r)=2\frac{dt}{T dr}
\end{equation}
The factor of 2 here accounts for the electron passing through a specific  $r$  twice per period.
Combining these  results, we obtain 
\begin{equation}
	\label{lneq0}
	p_{c}(r)=\frac{\sqrt{1+\frac{\ell^4}{{\tilde r}^2 \left[1-\frac{\ell^2}{n^2}-\left(1-\frac{\ell^2}{{\tilde r}}\right)^2\right]}}}{\pi n^3 a \sqrt{\frac{2}{{\tilde r}}-\frac{1}{n^2}}}\,.
\end{equation}
Interestingly,   this result does not reduce to Eq.(\ref{l20}), the result for $l=0$,  when we take 
the limit of $l\rightarrow 0$.   When $l\neq 0$, the classical orbit is elliptical 
and has two points where the radial velocity vanishes, leading to divergent radial probabilities. 
This is evident in the above equation, which has two singular points: 
$r_{1}/a=2\ell^2/(1-\epsilon)$ and $r_{2}/a=\ell^2/(1+\epsilon)$. These two singular points always exist
no matter how small the angular momentum is as long as it is not zero. In contrast, the probability 
density $p_{c}(r)$ has only one singular point for $l=0$.  

\subsection{Angular probability density}
The classical angular probability density $p_c(\theta)$ can be similarly defined as:
\begin{equation}
	\label{eq:ptheta}
	p_{c}(\theta)d\theta=\frac{dt}{T}=2\frac{ds}{vT}\,.
\end{equation}
The factor of 2 is included for the same reason as before. With Eq.(\ref{eq:rp}), 
it is clear that the relation in Eq.(\ref{eq:ptheta}) can be re-written as:  
\begin{equation}
	p_c(\theta)d\theta=2p(\gamma)d\gamma\,, 
\end{equation}
where $p(\gamma)$ is the probability density in terms of $\gamma$.  

To find $p(\gamma)$, we first consider the situation where $\vec{L}$ is fixed. In this case, 
as shown in Fig.\ref{fig:orbit}(b2), 
all the ellipses corresponding to a given eigenfunction $\psi_{nlm}(\vec{r})$ lie in the orbital plane $x'y'$ 
that is perpendicular to $\vec{L}$ , and these ellipses have different angle $\gamma_0$, which is 
indicated in Fig.\ref{fig:orbit}(b2)  and Fig.\ref{fig:ellipse}.  As we  average over all these ellipses, 
we will clearly get a uniform distribution, 
that is, $p(\gamma)=1/(2\pi)$. As this result is independent of the direction $\vec{L}$, we will get 
the same result when we average all the ellipses with different $\vec{L}$, that is, 
\begin{equation}
	\label{eq:gmth}
	p_c(\theta)d\theta=\frac{1}{\pi}d\gamma\,. 
\end{equation}
The remaining task is to find $d\gamma/d\theta$. 

We use Figure \ref{fig:ellipse}(b), where three additional lines AB, AC, and BC are drawn, 
to find the relation between $\theta$ and $\gamma$. These lines are drawn so that 
$\Delta$ABC is perpendicular to  the $Z$ axis. This immediately 
implies that OC$\perp$AC. Therefore, we have  $d_{\rm OC}=d_{\rm OA}\cos\theta$，
where $d_{\rm OC}$ and $d_{\rm OA}$ denote the lengths of OC and OA, respectively. 

On the other hand, as AB perpendicular to both the $Z$ axis and $\vec{L}$, 
AB is also perpendicular to the $X'$ axis. This gives us another relation between 
$d_{\rm OC}$ and $d_{\rm OA}$: $d_{\rm OC}=d_{\rm OB}\sin\alpha
=d_{\rm OA}\cos\gamma \sin\alpha$. Combining the two relations,  we eventually have 
\begin{equation}
	\label{eq:oc}
	\cos\theta=\cos\gamma \sin\alpha\,,  
\end{equation}
a relation between $\gamma$ and $\theta$.  As $\alpha$ is 
the angle between the angular momentum vector $\vec{L}$ and the $z$-axis, we have
\begin{equation}
	\cos\alpha=\frac{L_z}{L}=\frac{m}{\ell}\,.
\end{equation}
At the end we obtain
\be
\label{oc}
d\gamma=\frac{\sin\theta}{\sqrt{\sin^2\theta-\frac{m^2}{\ell^2}}}d\theta\,.
\ee
With Eq.(\ref{eq:gmth}), this leads to the classical angular probability density 
\begin{equation}
	\label{apc}
	p_c(\theta)=\frac{\sin\theta}{\pi \sqrt{\sin^2\theta-m^2/\ell^2}}
	=\frac{\sin\theta}{\pi \sqrt{\sin^2\theta-\cos^2\alpha}}\,,
\end{equation}
which diverges at two points $\theta=\pi/2\pm\alpha$.

We note that the derivation leading to the above angular probability density is independent of 
the form of the interaction $V(r)$ between the particle and the center.  Our argument that
$p(\gamma)=1/(2\pi)$ have not used any property of ellipse. In deriving Eq.(\ref{oc}), 
no properties of ellipse are used, either. In other words, we would get 
the same angular distribution even if the interaction between electron and proton is harmonic, i.e., 
$V(r)\propto r^2$. This is of course consistent with the result in quantum mechanics, 
where the angular wavefunction $Y_l^m(\theta,\phi)$ is independent of the form of $V(r)$. 

\section{Comparison between quantum and classical results}
In the semiclassical limit  $n\rightarrow \infty$,  we expect that the quantum  probability densities
converge to  their classical counterparts.  However, this comparison requires careful consideration 
of how the limit is taken. The classical probability densities are derived from an ensemble 
of orbits sharing identical eccentricity eccentricity $\epsilon$  and inclination $\alpha$. 
For a given eigenfunction  $\psi_{nlm}(\vec{r})$, the eccentricity $\epsilon$  is given 
by Eq.(\ref{eccen}) and the inclination is determined  
by $\cos\alpha=m/\ell$ (see Figure \ref{fig:ellipse}). 
To maintain physical consistency when taking $n\rightarrow \infty$, 
we must preserve both $\epsilon$  and $\alpha$ across  eigenfunctions with different $n$. 
This requires: keeping the ratio $l/n$ constant (fixing eccentricity) and maintaining 
$m/l$ constant for $l\neq 0$ (fixing inclination). For example, with fixed ratios $l/n=1/2$ and $l/m=1/2$, 
we examine and compare the sequence of  eigenfunctions  $\psi_{n,n/2,n/4}(\vec{r})$ as $n$ increases. 
This constrained limiting procedure ensures meaningful quantum-classical comparison and 
guarantees that the quantum probability densities properly converge to the classical ensemble results.

\subsection{The radial probability}
\begin{figure}[ht]
	\centering
	\begin{subfigure}[t]{0.3\textwidth}
		\centering
		\includegraphics[width=1\textwidth]{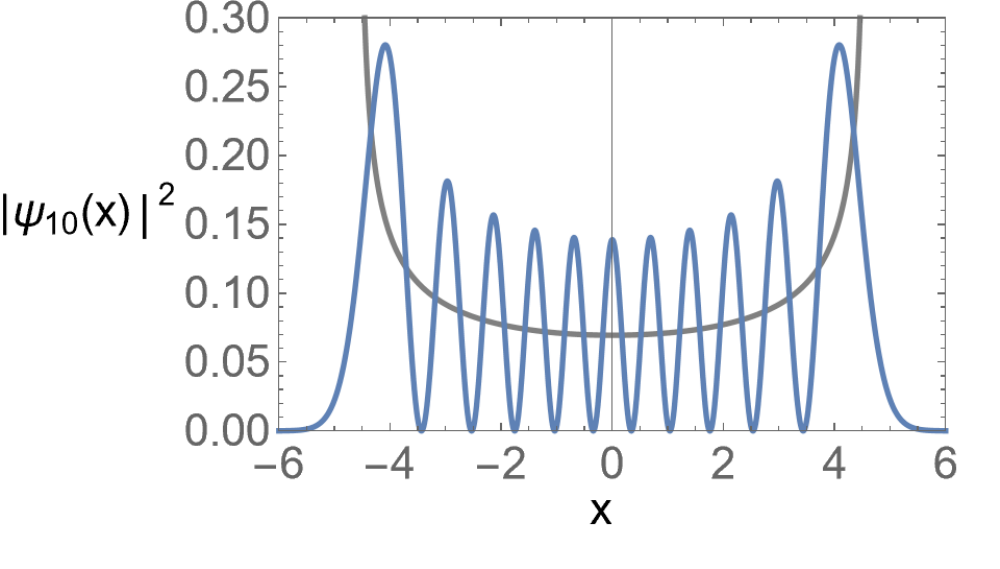}
		\subcaption{$n=10$, $l=0$}
	\end{subfigure}
	\quad
	\begin{subfigure}[t]{0.3\textwidth}
		\centering
		\includegraphics[width=1\textwidth]{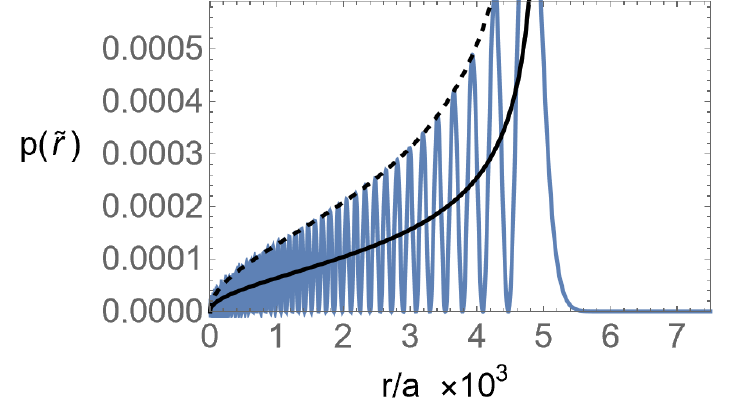}
		\subcaption{$n=50$, $l=0$}
	\end{subfigure}
	\quad
	\begin{subfigure}[t]{0.3\textwidth}
		\centering
		\includegraphics[width=1\textwidth]{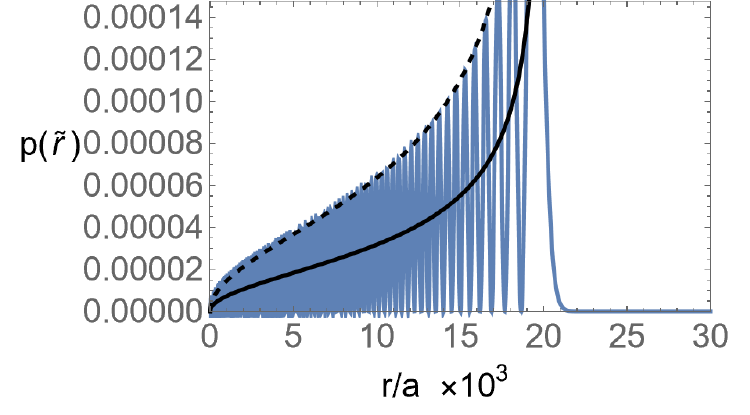}
		\subcaption{$n=100$, $l=0$}
	\end{subfigure}
	\caption{The radial probability density of the hydrogen atom 
		for the case of zero angular momentum ($l=0$). The blue lines
		are quantum results $p_q({\tilde r})$, 
		the black solid lines are classical results $p_c({\tilde r})$, and the dashed lines
		are doubled  classical results $2p_c({\tilde r})$. }
	\label{n00}
\end{figure}

The quantum radial probability density  for hydrogen atom is given by $p_{q}(r)=|R_{nl}(r)|^2 r^2$. 
To obtain dimensionless probability densities, 
we re-define $p_{c}(r)$ and $p_{q}(r)$ as $p_{c}(\tilde r)=p_{c}(r)a$ and $p_q(\tilde r)=p_{q}(r) a$. 
Figure \ref{n00} compares the quantum and classical probability densities for the case $l=0$. 
While the quantum probability oscillates with $r$, its overall 
trend aligns with its classical counterpart,  and the agreement becomes better as the quantum number 
$n$ increases.  When we plot twice the classical probability density,  
$2p_c({\tilde r})$ (dashed lines), they almost perfectly envelop the quantum oscillations (blue lines). 
This indicates that the quantum probabilities 
oscillate around the classical results.  If we were to smooth out these oscillations, 
the quantum and classical results would match almost perfectly for large $n$. 

At $r_c=2n^2a$, the classical electron has zero velocity, leading to divergence at this point. 
However, since the classical electron cannot move beyond $r_c$,  
its probability density immediately drops to zero for $r>r_c$. 
This abrupt cutoff is well-represented by the quantum curves, which exhibit a sharp decline beyond  $r_c$. 
Moreover, this decline becomes steeper as $n$ increases. 	

\begin{figure}[ht]
	\centering
	\begin{subfigure}[t]{0.45\textwidth}
		\centering
		\includegraphics[width=1\textwidth]{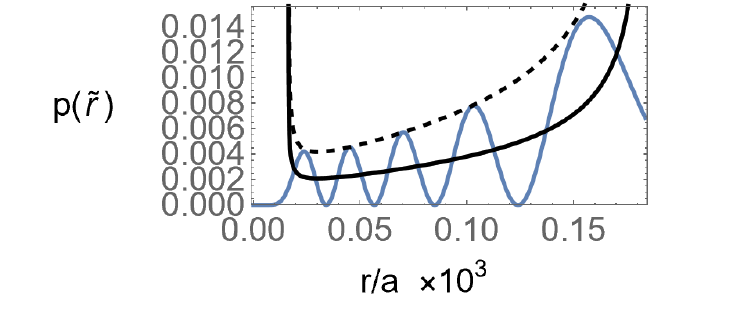}
		\subcaption{$n=10$, $l=5$}
	\end{subfigure}
	\quad
	\begin{subfigure}[t]{0.45\textwidth}
		\centering
		\includegraphics[width=1\textwidth]{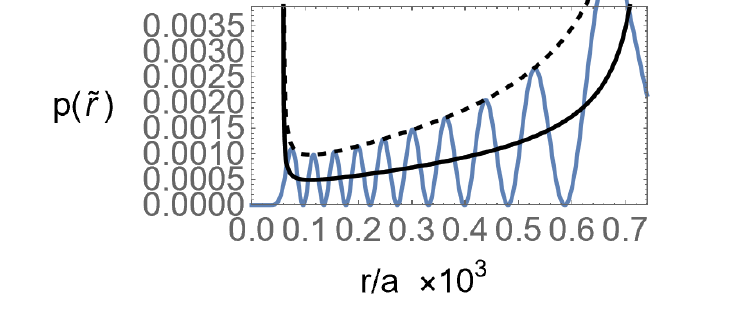}
		\subcaption{$n=20$, $l=10$}
	\end{subfigure}
	\begin{subfigure}[t]{0.45\textwidth}
		\centering
		\includegraphics[width=1\textwidth]{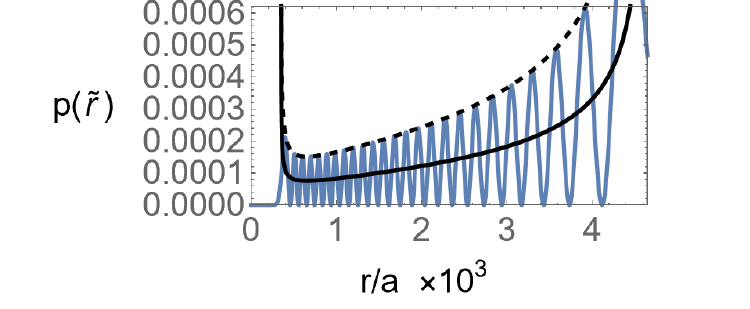}
		\subcaption{$n=50$, $l=25$}
	\end{subfigure}
	\quad
	\begin{subfigure}[t]{0.45\textwidth}
		\centering
		\includegraphics[width=1\textwidth]{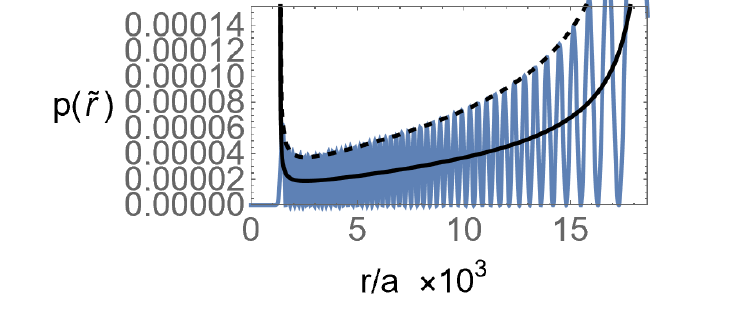}
		\subcaption{$n=100$, $l=50$}
	\end{subfigure}
	\caption{The radial probability density of hydrogen atom 
		for the case of non-zero angular momentum. The ratio $l/n$ is kept at $1/2$. 
		The blue lines are quantum results $p_q({\tilde r})$, 
		the black lines are classical results $p_c({\tilde r})$, and the dashed lines
		are doubled  classical results $2p_c({\tilde r})$.}
	\label{nl0}
\end{figure}

Figure \ref{nl0} presents a comparison of the radial probability densities 
for the case of non-zero angular momentum ($l\neq 0$). 
As we discussed at the beginning of this section, 
for a meaningful comparison, we must keep the ratio  $l/n$ 
constant as  $n$ becomes larger. As evident from 
Figure \ref{nl0},  there is a strong similarity between
the quantum and the classical results with the former 
oscillating around the latter. 

When the angular momentum is nonzero ($l\neq 0$), the classical orbit is elliptical 
and has two points where the radial velocity vanishes, leading to divergent radial probability. 
This is reflected by the fact that  Eq.(\ref{lneq0})  has two singular points at  
$r_{1}/a=2\ell^2/(1-\epsilon)$ and $r_{2}/a=\ell^2/(1+\epsilon)$. 
As shown in Figure \ref{nl0},  when $n$ is small, 
the classical diverging behavior at these two points, in particular, at $r_1$,  is not apparent 
at all in the quantum probability. However, as $n$ increases, the
quantum probability begins to develop sharp peaks around both $r_1$ and $r_2$, converging 
to the classical divergent behavior. 

The results presented in both Figure \ref{n00} and Figure  \ref{nl0} 
clearly demonstrate that the quantum radial probability 
density converges to its classical counterpart in the  limit of $n\rightarrow \infty$. 
These findings support our assertion that the quantum eigenfunction  $\psi_{nlm}(\vec{r})$ 
can be interpreted as an equal-weight superposition of classical orbits sharing the same energy 
and angular momentum. 

\subsection{The angular probability}
\begin{figure*}[htbp]
	\centering
	\begin{subfigure}[t]{0.45\textwidth}
		\centering
		\includegraphics[width=1\textwidth]{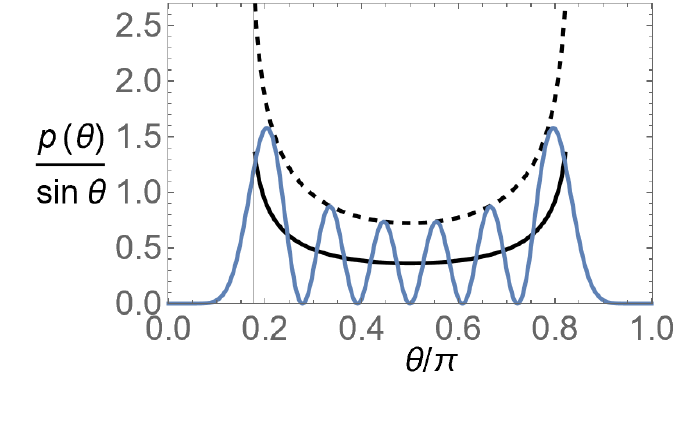}
		\subcaption{$l=10$, $m=5$}
	\end{subfigure}
	\quad
	\begin{subfigure}[t]{0.45\textwidth}
		\centering
		\includegraphics[width=1\textwidth]{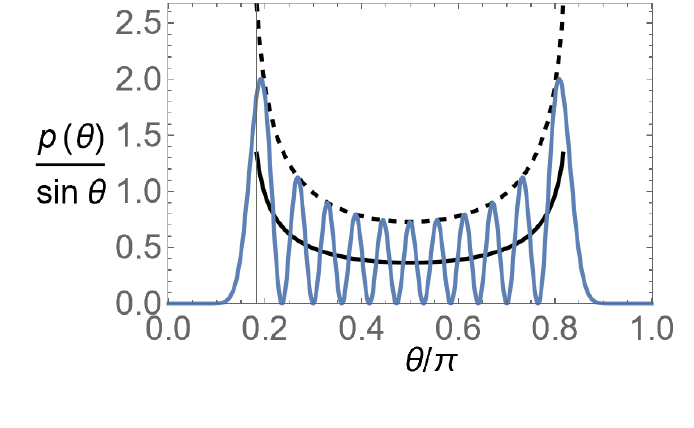}
		\subcaption{$l=20$, $m=10$}
	\end{subfigure}
	\begin{subfigure}[t]{0.45\textwidth}
		\centering
		\includegraphics[width=1\textwidth]{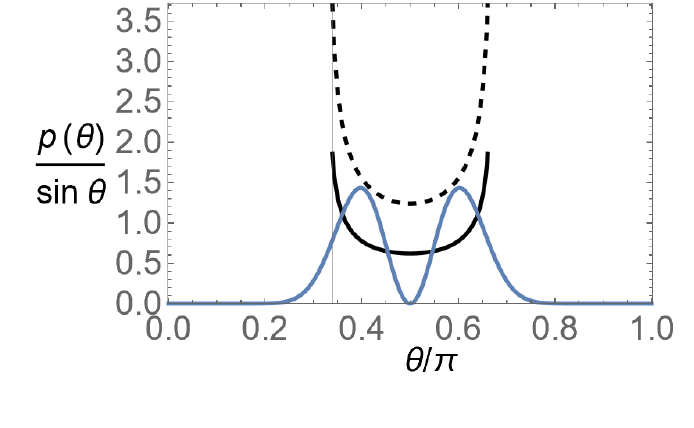}
		\subcaption{$l=10$, $m=9$}
	\end{subfigure}
	\quad
	\begin{subfigure}[t]{0.45\textwidth}
		\centering
		\includegraphics[width=1\textwidth]{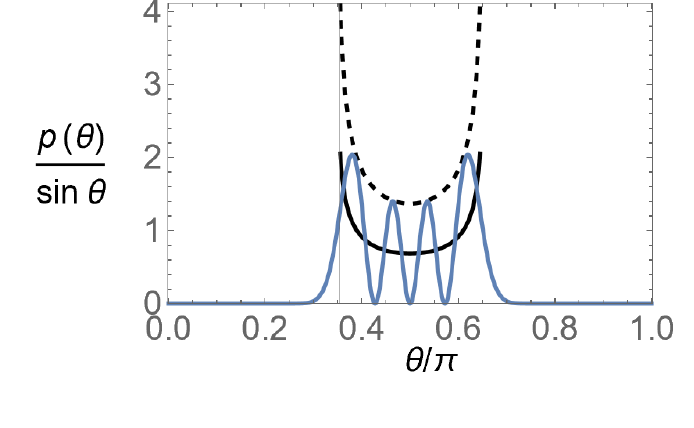}
		\subcaption{$l=30$, $m=27$}
	\end{subfigure}
	\caption{The angular probability density of hydrogen atom. 
		The ratios $l/n$ and $m/l$ are fixed.  The blue lines are quantum results $p_q(\theta)$,  
		the black solid lines are classical results $p_c(\theta)$, and the dashed lines are doubled  
		classical results $2p_c(\theta)$. }
	\label{nlm}
\end{figure*}

The quantum angular probability density for  hydrogen atom is given by:	
\begin{equation}
    \begin{aligned}
	p_{q}(\theta)
     &=|Y_l^m(\theta,\phi)|^2 \sin \theta\\
	   &=\frac{2l+1}{2} \frac{(l-m)!}{(l+m)!}\left[P_l^m(\cos\theta)\right]^2 \sin\theta\,,
    \end{aligned}
\end{equation}
which is independent of $\phi$. Its dependence on $L_z=m\hbar$ 
is implicit in the form of the associated Legendre function $P_l^m(\cos\theta)$. 
Eq.(\ref{apc}) reveals that 
the classical angular distribution $p_c(\theta)$ is also independent of $\phi$.
Figure \ref{nlm} compares the quantum and classical angular probability distributions. 
The figure shows that the trends of the two distributions are similar, 
and this similarity becomes more pronounced at higher $l$.

Together with the results for
the radial distribution, we  conclude that 	
the energy eigenfunction $\psi_{nlm}(\vec{r})$ of the hydrogen atom can be regarded 
as an equal-weight superposition of classical elliptic orbits with energy $E_n$, 
angular momentum $L=\sqrt{l(l+1)}\hbar$, and  $z$ component of angular momentum 
$L_z=m\hbar$. 

The quantum-classical correspondence that has been established is also evident in a qualitative manner. 
For both the quantum and classical models of the hydrogen atom, the radial probability 
density depends solely on quantum numbers \( n, l \) , with no dependence on the 
quantum number \( m \). Similarly, the angular probability 
density is a function only of \( l \) and \( m \), independent of \( n \).
The latter observation is particularly insightful because it implies that, 
although the shape of a single classical orbit is determined by the form of the central force \( V(r) \), 
the overall angular distribution of these classical orbits is independent of the specific form of \( V(r) \).

\section{Discussion and Summary}
We have calculated the classical probability densities for the hydrogen atom and compared 
them with their quantum counterparts. Our findings indicate that, in the semiclassical 
limit \( n \rightarrow \infty \), the energy eigenfunction \(\psi_{nlm}(\vec{r})\) of the hydrogen 
atom can be interpreted as a collection of classical elliptic orbits with energy \(E_n\), 
angular momentum \(L = \sqrt{l(l+1)}\hbar\), and \(z\)-component of angular momentum \(L_z = m\hbar\).

This serves as a special example of a broader question: What is the classical correspondence 
of an energy eigenfunction in the semiclassical limit? It is well established that in this limit, 
an energy eigenfunction reduces to an invariant distribution in classical phase 
space \cite{berry1,berry2,Zhang}. An ensemble distribution in classical phase space 
is considered invariant if it remains unchanged over time according to the Liouville equation. 
This general result can be seen directly via Moyal's equation\cite{Moyal}. 
For a given wave function $\psi(\vec{r},t)$, its corresponding Wigner function $W(\vec{p},\vec{r},t)$ satisfies Moyal's equation
\be
\label{moyal}
\begin{aligned}
\frac{\partial}{\partial t}W(\vec{p},\vec{r},t)&=\frac{2}{\hbar}
\sin\frac{\hbar}{2}\big[\frac{\partial}{\partial \vec{p}_W}\frac{\partial}{\partial \vec{r}_H}-
\frac{\partial}{\partial \vec{p}_H}\frac{\partial}{\partial \vec{r}_W}\big]\cdot\\
&\quad H(\vec{p},\vec{r})W(\vec{p},\vec{r},t)\,,
\end{aligned}
\ee
where  $\frac{\partial}{\partial \vec{p}_W},~\frac{\partial}{\partial \vec{r}_W}$ operate only on $W$
and $\frac{\partial}{\partial \vec{p}_H},~\frac{\partial}{\partial \vec{r}_H}$ only on $H$. 
It is evident that in the limit of $\hbar\rightarrow 0$, the Moyal's equation is reduced to Liouville equation 
and the Wigner function $W(\vec{p},\vec{r},t)$ becomes an ensemble distribution in classical phase space. 
In particular, when $\psi(\vec{r},t)$ is an energy eigenfunction, the corresponding Wigner function $W(\vec{p},\vec{r})$ 
is independent of time and becomes an invariant distribution in classical phase space. 
We define a classical trajectory $C$ as a collection of all the points that a particle 
traverses in phase space over an infinite long time for a given initial condition.    
An invariant ensemble must be made of various such trajectories. 

With this general result in mind, let us consider a central potential $V(r)$ other than the Coulomb potential. 
In this case,  the orbit may no longer be closed because the radial and angular motion have 
different periods\cite{unclose}. Nevertheless,  we expect a similar conclusion: the quantum eigenfunction is 
an equal-weight superposition of all possible classical orbits with the same energy $E$ 
and  the same angular momentum specified by $L$ and $L_z$. 
In particular, since the potential $V(r)$  is independent of $\theta$, 
the angular probability density remains the same and is described by Eq.(\ref{apc}). 
The radial probability density will be different from Eq.(\ref{l20}) but the classical radial density can 
be computed similarly as  $p_c(r)=2\frac{\sqrt{(dr)^2+(rd\gamma)^2}}{Tdr}$ with $T$ being 
the period of radial motion. 

We expect similar results for a scattering state ($E>0$): in the semiclassical limit, 
a scattering eigenstate of hydrogen can be viewed as a collection of many unbounded 
classical orbits with the same energy and the angular momentum. To study such a correspondence, for simplicity, one may first  
consider a one-dimensional scattering problem. Detailed study 
of this case is out of scope of this work.

In general, particularly in chaotic systems, the quantum energy eigenfunctions with eigen-energy $E_n$ 
can still be viewed as a collection of classical orbits with energy $E_n$. 
For hydrogen atom or other similar systems, each classical trajectory  contributes equally. 
However, it is  an open question 
how to determine or compute the weight of each classical orbit for a general system\cite{triangle}.

\begin{acknowledgments}
This work was supported by the National Natural Science Foundation of China (92365202, 12475011, 11921005), the National Key R\&D Program of China
(2024YFA1409002), the Shanghai Municipal Science and
Technology Major Project (2019SHZDZX01), the Shanghai
Municipal Science and Technology Project (25LZ2601100),
the Innovation Program for Quantum Science and Technology
(2021ZD0302100)
\end{acknowledgments}

\section*{Appendix A: A classical singularity problem from quantum perspective}
Since the 18th century, physicists and mathematicians have grappled with a 
simple question: What is the fate of a particle initially at rest falling 
toward a gravitational center? The divergence of the gravitational force 
at the origin presents a fundamental challenge to classical mechanics, 
as the equations of motion become singular at $r=0$\cite{Euler}. 

Euler addressed this problem by considering trajectories with small but 
finite angular momentum $L$ and taking the limit $L\rightarrow 0$.  
His analysis suggested the particle would execute a sudden reversal at the center. 
However, this conclusion sparked considerable debate among his contemporaries. 
Laplace, for instance, argued that there is a crucial distinction between  
the limit of flattened elliptical orbits  ($L\rightarrow 0$) and strictly radial motion. 
It is much like how a function's limiting behavior may differ from its value at a point. 
This perspective found support from D'Alembert, who argued for the possibility of the particle 
passing through the center. Some modern 
interpretations suggest the particle might halt at the center\cite{Euler}.  

Given the Schr\"odinger equation has a well-defined solution for a particle at rest falling into a
gravitational center, i.e., the case of zero angular momentum $l=0$,  a natural question is whether
the quantum-classical correspondence established in the main text could resolve this controversy. 
Unfortunately, according to the following analysis, the quantum-classical correspondence does not
resolve the issue but does provide some new insights.

\begin{figure}[ht]
	\centering
	\begin{subfigure}[t]{0.5\textwidth}
		\centering
		\includegraphics[width=1\textwidth]{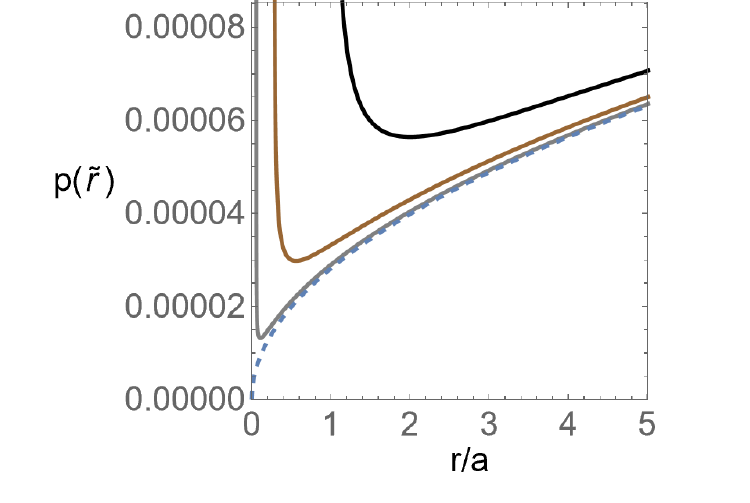}
		\subcaption{}
	\end{subfigure}
	\quad
	\begin{subfigure}[t]{0.4\textwidth}
		\centering
		\includegraphics[width=1\textwidth]{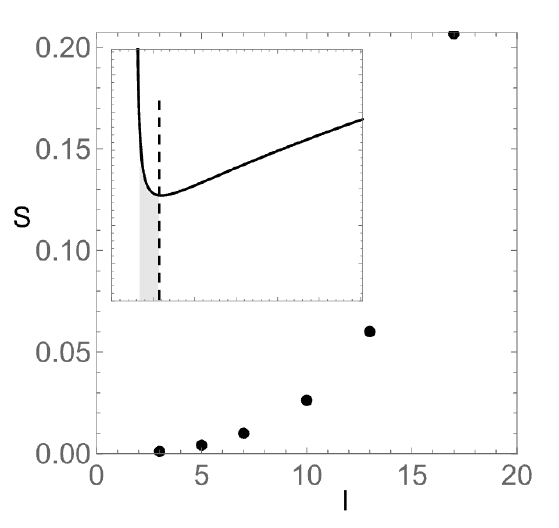}
		\subcaption{}
	\end{subfigure}
	\quad
	\caption{(a) The radial probability density of the classical hydrogen model for three different 
		values of $l=1$ (gray), $0.4$ (brown), $0.1$ (black) at $n=20$. For comparison, the density for $l=0$ is drawn as dashed line.  
		(b) The area of
		the divergent peak as a function of the angular momentum $l$ at $n=20$. 
		The inset shows how the area $S$ 
		is defined for a density function. The dashed line in the inset pass the turning point where 
		the derivative of the density function is zero.}
	\label{csl}
\end{figure}

Since the singularity occurs at $r=0$, 
we will proceed to examine how the radial distribution, 
classical or quantum,  behaves in the vicinity of $r=0$. 
We first look at the classical radial distribution. In the case of  zero angular momentum, i.e., $l=0$, 
the distribution is given by  Eq.(\ref{l20}), which becomes 
\begin{equation}
	\label{l200}
	p_{c}(r)\approx \frac{1}{\pi  n^3 a^{3/2}}\sqrt{\frac{r}{2}}\,,
\end{equation}
when $\tilde{r}\rightarrow 0$. It is well behaved. In contrast, the radial distribution 
for $l\neq 0$, given by Eq(\ref{lneq0}) ,  diverges at 
$r_{2}/a=\ell^2/(1+\epsilon)$.  Specifically, 
near $\tilde{r}=r_{2}/a$, the radial distribution for non-zero angular momentum 
exhibits the following behavior: 
\begin{equation}
	\label{l201}
	p_{c}(r)\approx \frac{\ell r_{2}}{2\pi  n^3 a^{3/2}}\sqrt{\frac{\epsilon}{r-r_{2}}}\,. 
\end{equation}
In the classical framework, the angular momentum $l$ is not necessary an integer and 
can take any positive real number. Consequently, we can consider the limit where $l$ approaches zero.  
As illustrated in Figure \ref{csl}(a), the divergent behavior persists in this limit. 
In fact, this divergent behavior of the distribution $p_{c}(r)$ is still described by Eq.(\ref{l201})  with $\ell$
being a small real number. 

The above results indicate that the classical radial distribution $p_{c}(r)$
for non-zero angular momentum ($l \neq 0$), as given by Eq. (\ref{lneq0}), 
does not converge to Eq. (\ref{l20}) in the limit where the angular momentum approaches 
zero ($l \rightarrow 0$). This observation suggests that there is indeed a fundamental distinction 
between motion along an infinitely flattened ellipse and motion along a straight line, thereby lending 
support to Laplace's assertion.  

However, from a different perspective, one might conclude that Euler was correct. Since we are 
dealing with a probability density function, it is essential to consider how the area 
under the divergent peak changes as \(l\rightarrow 0\). The area \(S\) under the divergent 
peak is defined in the inset of Figure \ref{csl}(b) and represents the area under the density function 
between the divergent point and the turning point. As shown in Figure \ref{csl}(b), our calculations indicate 
that the area \(S\) tends to zero as \(l\rightarrow 0\). This implies that, statistically, the peak of 
the distribution (\ref{lneq0}) near \(r=0\) can be ignored when \(l\) is very small. In this sense, 
the two density functions (\ref{l20}) and (\ref{lneq0}) are in agreement in the limit of \(l\rightarrow 0\). 
Therefore, we cannot definitively conclude that Laplace was right and Euler was wrong. 

Let us now examine whether the quantum radial distribution can provide any new insights 
into this singularity problem. As illustrated in Figure \ref{nl0}, the first peak of the 
quantum radial distribution \(p_q(r)=|R_{nl}(r)|^2\) diverges in the limit of \(l\rightarrow\infty\) 
and \(n\rightarrow\infty\) while keeping \(l/n\) fixed, mirroring the behavior of the classical radial distribution. However, when \(l\) is zero, no such divergent behavior is observed, as seen in Figure \ref{n00}.
Thus, similar to the classical case, the quantum radial distribution for \(l\neq 0\) does not 
approach that for \(l=0\) in the limit of \(l\rightarrow 0\).

\begin{figure}[ht]
	\centering
	\includegraphics[width=0.4\textwidth]{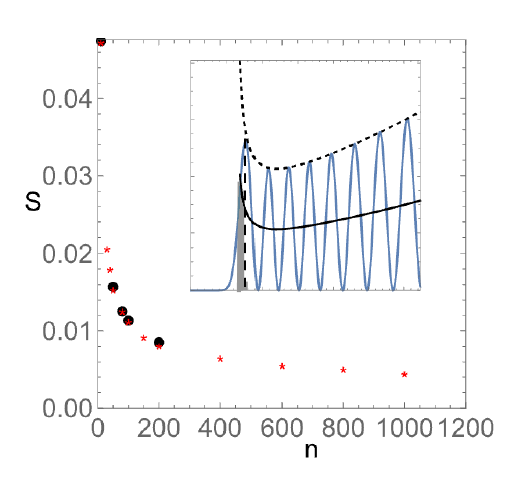}
	
	\caption{The quantum peak area $S$ as a function of $n$. The corresponding classical 
		area is defined in the inset. The quantum results are represented as circles and the classical ones as 
		stars. The computations are done with $l/n=1/2$ fixed. }
	\label{q}
\end{figure}

To be precise, we need to consider the weight of the diverging peak. For the quantum 
probability density \(p_q(r)=|R_{nl}(r)|^2\), the weight of the peak is defined as the area 
under the function between \(r=0\) and the first peak. The corresponding classical area is defined 
as the area between the divergent point and the first quantum peak, as shown in the inset of Figure \ref{q}.
Note that this classical area is different from the one defined in the inset of Figure \ref{csl}; 
however, they are proportional to each other.

The computed quantum peak areas are shown as a function of \(n\) in Figure \ref{q}, 
along with the classical area. The quantum peak becomes increasingly difficult 
to compute as \(n\) becomes very large; as a result, we have computed only up to \(n=200\). 
However, as seen in Figure \ref{q}, the quantum peak area is almost identical to the corresponding 
classical area. Therefore, we have computed the classical areas for larger \(n\), and the results show 
that the peak area tends to zero as \(n\rightarrow\infty\).
This indicates that the diverging peak near \(r=0\) in the quantum radial distribution can be statistically ignored.

In conclusion, the singularity problem of a particle initially at rest falling toward a gravitational center 
cannot be definitively resolved using distribution functions. On one hand, the diverging behavior 
of the radial distributions at the limit of \(l/n\rightarrow 0\) with \(n,l\rightarrow\infty\)
suggests that the particle would make a sudden turn through the gravitational center, 
supporting Euler's view. However, as the weight of the divergent peak tends to zero, 
one cannot be entirely certain and rule out Laplace’s suggestion that 
the particle would pass through the center. 
Nevertheless, we can rule out a recent  suggestion that  the particle might halt at the center\cite{Euler}. 
If the particle  halt at the center, we would have a peak of non-zero weight near $r=0$ 
for  distribution functions. 

While our analysis 
provides some fresh insights, it cannot fully resolve the singularity issue. Our findings 
indicate that (1) the particle does not remain permanently at the center and (2) 
the question of whether Euler's turning-point solution is physically valid remains open.
This historical controversy highlights the subtle relationship between limiting processes 
and physical reality - a theme that recurs throughout mathematical physics.

\section*{Data availability statement}
This study does not have experiment data.

\end{document}